\documentclass[fleqn,10pt]{Notes} % Document font size and equations flushed left

\definecolor{color1}{RGB}{0,0,90} % Color of the article title and sections

\newlength{\tocsep} 
\usepackage{amssymb}
\usepackage{amsmath}
\usepackage{enumerate}
\usepackage{enumitem}
\usepackage{times}
\usepackage{mathtools}
\usepackage{natbib}

\DeclareMathAlphabet{\bi}{OML}{cmm}{b}{it}

\newcommand{\vv}[1]{\mathbf{#1}}

\JournalInfo{2019} % Journal information
\Archive{Research article (preprint)} % Additional notes (e.g. copyright, DOI, review/research article)

\PaperTitle{Collapse and recovery times in non-linear harvesting with demographic stochasticity
}

\Authors{Sara Cuenda\textsuperscript{1,2}, Marta Llorente\textsuperscript{1,2}, José A. Capitán\textsuperscript{3}*} 
\affiliation{\textsuperscript{1}\textit{Universidad Aut\'onoma de Madrid. Depto. An\'alisis Econ\'omico: Econom{\'\i}a Cuantitativa. C/ Francisco Tom\'as y Valiente 5, 28049 Madrid, Spain}} 
\affiliation{\textsuperscript{2}\textit{Grupo de Economía Matemática Aplicada, Universidad Complutense de Madrid}
}
\affiliation{\textsuperscript{3}\textit{Complex systems group. Department of Applied Mathematics. 
Universidad Polit\'ecnica de Madrid. Av. Juan de Herrera, 6. 28040 Madrid, Spain}}
\affiliation{*\textbf{Corresponding author}: ja.capitan@upm.es}

\Keywords{Non-linear harvesting --- Bifurcation theory --- Demographic stochasticity --- Continuous-time Markov process --- Mean
first-passage time}
\newcommand{\keywordname}{Keywords} % Defines the keywords heading name

\Abstract{
Recent collapses of many fisheries across the globe have challenged the mathematical approach to these systems through classic bioeconomic 
models. Decimated populations did not recover as fast as predicted by these models and depensatory effects were introduced to better fit the dynamics
at low population abundances. Alternative to depensation, modeling captures by non-linear harvesting functions produces equivalent outcomes at small 
abundances, and the dynamics undergoes a bifurcation leading to population collapse and recovery once catching efforts are above or below certain 
thresholds, respectively. The time that a population takes to undergo these transitions has been mostly overlooked in bioeconomic contexts, though. In 
this work we quantify analytically and numerically the times associated to these collapse and recovery transitions in a model incorporating non-linear 
harvesting and immigration in the presence and absence of demographic stochasticity. Counterintuitively, although species at low abundances are prone 
to extinction due to demographic stochasticity, our results show that stochastic collapse and recovery times are upper bounded by their deterministic 
estimates. This occurs over the full range of immigration rates. Our work may have relevant quantitative implications in the context of fishery 
management and rebuilding. 
}

\begin{document}

\flushbottom % Makes all text pages the same height
\maketitle % Print the title and abstract box
%\tableofcontents % Print the contents section
\thispagestyle{empty} % Removes page numbering from the first page

%----------------------------------------------------------------------------------------
%	ARTICLE CONTENTS
%----------------------------------------------------------------------------------------

\section{Introduction}
\label{sec:intro}

Classic bioeconomic models have been extensively questioned after the recent dramatic collapse of many fisheries. It is estimated that marine fish abundance has declined 38\% globally since 1970 and that the rate of decline among top predators has increased since 1992~\citep{hutchings:2011}. Many fish stocks worldwide have been overexploited to the point of commercial extinction~\citep{fao:2012,simms:2017} and some have failed to recover despite the imposition of severe catch restrictions including moratoriums. Several studies~\citep{myers:2003} show the heavy overexploitation of all of the commercially important fisheries in the world. As of today, most populations exhibit little to moderate changes in abundance after fifteen years from their collapse~\citep{rose:2015}. 

Among depleted stocks, the collapse of Northern cod fishery stands as a paradigmatic social and economic disaster and almost certainly represented the greatest numerical loss of a vertebrate species in Canada~\citep{hutchings:2011}. Atlantic cod was the second most important species in 1970 with 3.1 million tonnes and only the sixth most abundant species in 1989, and the tenth most important species in 1992~\citep{hutchings:2011}. After decades of severe overexploitation, the northern cod stock collapsed to an extremely low level in 1992. A moratorium was imposed on the commercial fishery for northern cod in July 1992, but overfishing was sufficiently high to cause this population a collapse. Fisheries were re-opened in the inshore in 2006 and continued from 2007 on. More than two decades after the moratorium, the decimated northern cod stock shows signs of a slow rebuilding but still remains at a low abundances relative to those of 1980.

Despite many factors contributing to the overall fishing mortality such as environmental factors, poor recruitment to the fishery, or misreported catches of young individuals, there is a general agreement that fish stocks have been overexploited and that collapses are not consistent with some of the classical bioeconomic models hypotheses. In the standard biomass deterministic model by \cite{schaefer:1957}, populations are modeled using a logistic law of 
growth~\citep{verhulst:1838}, $F(y)=ry\left(1-\frac{y}{K}\right)$, where $r$ is the intrinsic growth rate and $K$ is the carrying capacity of the environment. Such population is harvested at a rate proportional to the total abundance $y$ through a production function $H(y)=cy$, where $H$ represents the total catch per unit time and the harvesting rate, $c:=qE,$ is constant with $E$ denoting fishing effort (the combined flow of labor and produced capital services) and $q$ the catchability. With these assumptions, the population dynamics for the resource is given by
\begin{equation}\label{eq:schaefer}
\frac{dy}{dt}=F(y)-H(y)=ry\left(1-\frac{y}{K}\right)-cy.
\end{equation}
The standard biomass approach uses conventional regression methods to find the curve of best fit for the stock data among the given family of logistic curves~\citep{gordon:1954,schaefer:1957,clark:1990}. Being these curves concave even at low levels of stock, this approach assumes the a priory hypothesis of compensatory dynamics at low population sizes. Regardless of the apparent prevalence of compensatory population dynamics [c.f. Eq.~\eqref{eq:schaefer}], the great resilience on the species assumed by concavity hypothesis does not agree with the observed collapses and biological reference points used in conventional fishery management are likely to be inaccurate and possibly nonconservative~\citep{maroto:2014}.

The observed lack of population recovery after a dramatic population reduction, even when fishing catches had been reduced, brought the attention to study population dynamics at small population sizes, leading to models supporting depensation or critical depensation~\citep{dennis:1989,hutchings:1996,alvarez:1998,maroto:2014}, for which the net growth rate $y'(t)$ is a convex function of $y$ at small abundances. These models fit better with the observed lack of recovery as per capita growth rate becomes negative for low enough stock values producing an unstable equilibrium at low population size, a critical population abundance below which extinction is certain. \cite{maroto:2014} introduced a methodological approach using splines at low abundances to remove the concavity hypothesis. The authors analyzed the stochastic population dynamics of Northern Cod stocks at low stock levels to provide management strategies ensuring long term sustainability. They showed that the combined effects of demographic uncertainty and depensation at low population sizes agreed with the observed collapse for this species.

Regarding the production function $H(y)$, the  catch-per-unit-effort hypothesis in \eqref{eq:schaefer} has also been questioned for lacking realism. On the one hand, \cite{clark:1979} observed that the implicit assumption of random (and independent) search for uniformly distributed fish is often violated in actual fisheries. On the other hand, it assumes a catch-rate that increases linearly as the population size gets larger and larger for a fixed level of effort. The inverse effect of fish abundance on the fishing effort is then not taken into account:  a greater harvest would induce a lower price of harvest which, in turn, induces less fishing effort. If we assume that the market price of the harvest motivates changes in fishing effort, large population levels should induce less fishing efforts~\citep{idels:2008}.

An alternative to introduce depensation in population dynamics~\eqref{eq:schaefer} is to replace the production function $H(y)$ by a non-linear function of abundances~\citep{clark:1979}. The resulting model leads (under suitable parameter combinations) to ``catastrophic'' fishery collapses as effort $E$ or abundance increases, producing a similar effect to the Schaefer catch equation if the growth function  is convex at small abundances~\citep{clark:1990}. The model proposed by~\cite{clark:1979} assumes a non-linear production function [also know as Holling type II functional response, see~\cite{holling:1959}], which saturates at large population abundances due to limitations in harvesting efforts. As derived in subsequent contributions~\citep{jones:1976,gulland:1977,agnew:1979,evans:1981,chaudhuri:1990}, this model exhibits a range in harvesting rates for which three critical points coexist, due to convexity at low abundances, leading to bi-stability and to transitions between two stable states. Above an upper threshold in harvesting, a single critical point remains, resulting in a pitchfork-type bifurcation. Close to the limits of the bi-stability range, transitions between the two stable attractors are known to occur, and little is known about the time behavior of these transitions, which are ultimately related to the collapse and the recovery of the population.

In this work we analytically and numerically quantify the times associated to these collapse and recovery transitions in a population model which incorporates non-linear harvesting. To make the setting more general, we study population dynamics in the presence of immigration, which introduces individuals in the system a constant rate (not necessarily small). We provide analytical estimates for deterministic collapse and recovery times, and incorporate demographic stochasticity to include the effect of variability associated to small population sizes, which is known to affect population dynamics in non-trivial ways~\citep{capitan:2015,capitan:2017}. We apply the formalism of mean first-passage time to define consistently how to measure collapse and recovery times in the presence of stochasticity. Counterintuitively, although individual deaths are increased at low population numbers due to stochasticity, our main result shows that stochastic collapse and recovery times are smaller than their deterministic estimates. This occurs over the full range of immigration rates. We also show that, in the limit of large carrying capacity, stochastic times actually converge to deterministic predictions. We finally discuss the implications of our work in the context of fishery management and rebuilding. 

\section{Deterministic model}
\label{sec:deterministic}

Consider a single population whose dynamics in isolation is determined by a logistic growth, defined by a carrying capacity $K$ and an intrinsic growth 
rate $r$. The system is not closed, and the population is externally harvested through individual catches, which are modeled by a non-linear functional 
response of the population abundance $y$, $f(y)=\frac{cy}{1+by}$, with saturation (due to limitations in catching efforts, captures remain finite for large
population levels). Parameters $b$ and $c$ determine the shape of the harvest rate, such that in the limit $b\to 0$ captures are a linear function of the 
population abundance $y$, as routinely assumed in bioeconomic models~\citep{schaefer:1957}, see Eq.~\eqref{eq:schaefer}. In addition, new individuals can arrive to the system at a constant 
immigration rate $I$. The dynamics is driven by the following differential equation:
\begin{equation}\label{eq:immi}
\frac{dy}{dt}=ry\left(1-\frac{y}{K}\right)-\frac{cy}{1+by}+I.
\end{equation}
The dynamics can be re-scaled by changing to a non-dimensional temporal variable $\tau:=rt$. In addition, we re-scale population abundance as $x:=by$. 
Then Eq.~\eqref{eq:immi} reduces to
\begin{equation}\label{eq:imscal}
\frac{dx}{d\tau}=x\left(1-\frac{x}{\sigma}\right)-\frac{\rho x}{1+x}+\mu,
\end{equation}
where new model parameters $\sigma:=Kb$ (scaled carrying capacity), $\rho:=c/r$ (scaled catching effort) and $\mu:=bI/r$ (scaled immigration) 
are non-dimensional.

\begin{figure*}[t!]
\begin{center}
\hspace*{-5mm}
\includegraphics[width=1.02\textwidth]{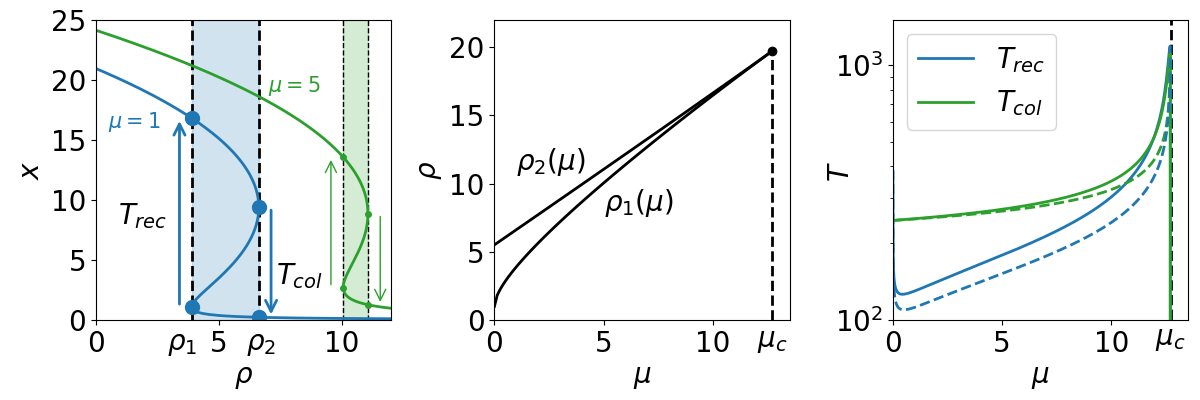}
\caption{
\label{fig:deterministic} 
\emph{Deterministic collapse and recovery times}. The left panel shows the solutions of Eq.~\eqref{eq:cubic} in terms of the harvesting rate $\rho$, 
where three-equilibria coexistence regions are shadowed. The transitions associated to collapse and recovery times are marked with arrows. The middle 
panel depicts model's parameter space, showing the thresholds $\rho_{1,2}$ as functions of immigration, see Eq.~\eqref{eq:rho12}. In the right panel, 
collapse and recovery times have been plotted in terms of $\mu$ according to Eqs.~\eqref{eq:collapse} and~\eqref{eq:recovery}. Dashed lines correspond
to approximations~\eqref{eq:colapp} and~\eqref{eq:recapp}. For $\mu\ge \mu_c$ the system is fully reversible and, as a consequence, we set 
$T_{col}=T_{rec}=0$. Here we used $\Delta\rho=0.001$ and $\Delta x=0.01$. In all the panels we set $\sigma=20$.
}
\end{center}
\end{figure*}

The non-linear harvesting term for captures introduces a pitchfork-type bifurcation in the absence of immigration~\citep{clark:1990}. Indeed, the same 
holds for non-zero immigration rates. The equilibrium abundance satisfies a cubic equation,
\begin{equation}\label{eq:cubic}
x(1+x)\left(1-\frac{x}{\sigma}\right)-\rho x+\mu(1+x)=0.
\end{equation}
Depending on the stability of its non-negative solutions, the dynamics may exhibit a single, stable equilibrium point or, within a range of parameters,
two stable equilibria coexist with an unstable point ---it is straightforward to see that the unstable rest point is located in between of the two
stable ones. In particular, there is an interval $(\rho_1,\rho_2)$ of catching efforts over which three equilibrium points coexist (see 
Fig.~\ref{fig:deterministic}, left). That range can be obtained by solving Eq.~\eqref{eq:cubic} for $\rho$ in terms of $x$,
\begin{equation}\label{eq:rhox}
\rho(x)=(1+x)\left(1-\frac{x}{\sigma}+\frac{\mu}{x}\right),
\end{equation}
a function defined over the interval $x\in[0,\infty)$. For non-negative abundances, this function has a relative maximum located at $(x_2,\rho_2)$ 
and a minimum at $(x_1,\rho_1)$, see Fig.~\ref{fig:deterministic} (left). The range of $\rho$ where bi-stability arises is precisely the interval 
$(\rho_1,\rho_2)$. The condition
for $x_{1,2}$ is simply $\rho'(x)=0$, i.e.,
\begin{equation}\label{eq:rhop}
x^2(\sigma-1-2x)=\sigma\mu.
\end{equation}
Figure~\ref{fig:deterministic} (middle panel) shows that the range $(\rho_1,\rho_2)$ shrinks as immigration increases. There is a maximum immigration rate 
$\mu_c$ below which this range in $\rho$ emerges. Such threshold in immigration can be calculated as follows: let $\mu_c$ be such that the abscissae of 
the maximum and minimum coincide, $x_1=x_2=:x_c$, i.e., $\rho(x)$ has an inflection point with zero slope at $x=x_c$. Then $x_c$ satisfies the conditions 
$\rho'(x_c)=0$ and $\rho''(x_c)=0$, which reduce to $x_c^2(\sigma-1-2x_c)=\sigma\mu_c$ and $x_c^3 = \sigma\mu_c$, respectively. Solving for $x_c$ 
and $\mu_c$ yields the threshold in immigration
\begin{equation}\label{eq:muc}
\mu_c:=\frac{1}{\sigma}\left(\frac{\sigma-1}{3}\right)^3
\end{equation}
in terms of the scaled carrying capacity. For $\mu<\mu_c$, within the interval $(\rho_1,\rho_2)$ three equilibria coexist due to the folding 
of the curve $x=x(\rho)$. For $\mu\ge\mu_c$, a single, non-negative rest point remains.

Analytical expressions for threshold values $\rho_{1,2}$ that enclose bi-stability can be found: at those limiting values of the harvesting rate, 
equilibrium abundances $x_1$ and $x_2$ are expressed in terms of $\mu$ and $\sigma$ as
\begin{equation}\label{eq:x12}
\begin{aligned}
&x_1(\mu,\sigma)= \frac{\sigma-1}{6}\left[1 - 2\sin\left(\frac{1}{3}\sin^{-1}\left(1 - \frac{2\mu}{\mu_c}\right)\right)\right],\\
&x_2(\mu,\sigma)= \frac{\sigma-1}{6}\left[1 + 2\cos\left(\frac{1}{3}\cos^{-1}\left(1 - \frac{2\mu}{\mu_c}\right)\right)\right],
\end{aligned}
\end{equation}
respectively (these formulae are derived in Appendix~\ref{sec:appA}). Now from condition~\eqref{eq:rhop} we can substitute 
$\mu=x_{1,2}^2(\sigma-1-2x_{1,2})/\sigma$ into Eq.~\eqref{eq:rhox}, so the limits $\rho_{1,2}$ to the harvesting rate are written as
\begin{equation}\label{eq:rho12}
\rho_{1,2}=\left(1+x_{1,2}\right)^2\left(1-\frac{2x_{1,2}}{\sigma}\right).
\end{equation}
Substituting expressions~\eqref{eq:x12} into Eq.~\eqref{eq:rho12} yields analytical functions for the way threshold values $\rho_{1,2}$ depend on 
immigration rate $\mu$ and scaled carrying capacity $\sigma$. These limits are plotted in Fig.~\ref{fig:deterministic} (middle panel) as a function of 
$\mu$. As immigration increases, two combined effects arise: (i) the range of bi-stability shrinks, and (ii) the bi-stability range emerges for larger values of 
catching efforts.

%\begin{equation}
%\begin{aligned}
%&\rho_1= \frac{1}{\sigma}\left[\frac{\sigma+5}{6}-\frac{2(\sigma-1)}{6}\sin\left(\frac{1}{3}\sin^{-1}\left(1 - \frac{2\mu}{\mu_c}\right)\right)\right]^2
%\left[\frac{2\sigma+1}{3}+\frac{2(\sigma-1)}{3}\sin\left(\frac{1}{3}\sin^{-1}\left(1 - \frac{2\mu}{\mu_c}\right)\right)\right],\\
%&\rho_2= \frac{1}{\sigma}\left[\frac{\sigma+5}{6}+\frac{2(\sigma-1)}{6}\cos\left(\frac{1}{3}\cos^{-1}\left(1 - \frac{2\mu}{\mu_c}\right)\right)\right]^2
%\left[\frac{2\sigma+1}{3}-\frac{2(\sigma-1)}{3}\cos\left(\frac{1}{3}\cos^{-1}\left(1 - \frac{2\mu}{\mu_c}\right)\right)\right].
%\end{aligned}
%\end{equation}

\subsection{Estimates to collapse and recovery times}
\label{ssec:times}

Let us consider the effect of varying the harvesting rate $\rho$ in a hysteresis cycle. Assume that the system is at equilibrium for $\rho<\rho_1$. In this 
region there is a single non-negative rest point with abundance far away from zero. With this initial condition, a small increase in $\rho$ will move the 
system to the corresponding equilibrium point following the curve $x=x(\rho)$ depicted in Fig.~\ref{fig:deterministic} (left panel). Now augment $\rho$ 
adiabatically, i.e., letting the system to relax to equilibrium before the next increase in the harvesting rate. Because perturbations are small (we assume
that the initial condition always belongs to the basin of attraction of the corresponding equilibrium point), even in the region where three rest points 
coexist, the population moves along the upper branch of the curve until $\rho=\rho_2$. Once this threshold is crossed over, a single 
stable point close to extinction remains, and the population \emph{collapses} to this point. The system undergoes a transition that cannot restore the 
initial abundance by simply adiabatically decreasing the harvesting rate, because the population now is in the lower branch of the curve. Small subsequent
diminutions move the system to larger abundances but close to extinction, and the system does not \emph{recover} its original state until $\rho<\rho_1$.
In this subsection we provide sensible estimates for the time spent in collapse (once $\rho$ is immediately above the threshold $\rho_2$) and to
recover abundances far from extinction (once $\rho$ is immediately below the threshold $\rho_1$).

In both cases the initial population abundance is equal to the equilibrium abundance given by Eq.~\eqref{eq:x12}: (i) $x=x_2$ for collapse
situations and (ii) $x=x_1$ in a recovery scenario. Then the system is perturbed by (i) augmenting or (ii) lowering the harvesting rate $\rho$. In both cases, 
there is a single real solution of the cubic equation~\eqref{eq:cubic}, the two remaining roots being complex conjugate. We can re-write the condition 
for equilibrium abundances as
\begin{multline}\label{eq:complex}
x(x+1)(x-\sigma)+\sigma\rho x-\sigma\mu(x+1)\\
  =(x-x_e)\left[(x-\gamma)^2+\beta^2\right]=0,
\end{multline}
where $x_e$ is the real root of the cubic polynomial and $\gamma$, $\pm\beta\in \mathbb{R}$ are the real and imaginary parts of the complex roots. 
The differential equation~\eqref{eq:imscal} can be written as
\begin{equation}
\frac{dx}{d\tau}=-\frac{(x-x_e)\left[(x-\gamma)^2+\beta^2\right]}{\sigma(x+1)},
\end{equation}
which can be immediately integrated to yield 
\begin{equation}
\tau(x;\vv{p}):=-\sigma \int \frac{x+1}{(x-x_e)\left[(x-\gamma)^2+\beta^2\right]}dx.
\end{equation}
Here $\vv{p}=(\rho,\mu,\sigma)$ stands for the vector of model parameters; note that $\tau$ implicitly depends on $\vv{p}=(\rho,\mu,\sigma)$ through 
$x_e$, $\gamma$ and $\beta$ [see Eq.~\eqref{eq:complex}]. An explicit expression of a primitive function is
\begin{multline}\label{eq:T}
\tau(x;\vv{p})=
\sigma\left[\frac{(\gamma+1)(x_e - \gamma) - \beta^2}{\beta[(x_e-\gamma)^2 + \beta^2]}\tan^{-1}\left(\frac{x-\gamma}{\beta}\right)\right.\\
\left.-\frac{1+x_e}{2[(x_e-\gamma)^2 + \beta^2]}\log\frac{(x - x_e)^2}{(x-\gamma)^2 + \beta^2}\right].
\end{multline}
We can obtain the sought times to collapse and recovery as follows:
\begin{itemize}
\item[(i)] \emph{Collapse time}. Starting from the initial condition $x=x_2$, catches mildly increase from $\rho_2$ to $\rho_2'=\rho_2+\Delta\rho$ 
with $\Delta\rho$ a small increment. The time to fall to the low-abundance endpoint $x_e$ (obtained numerically as the only real solution of 
Eq.~\eqref{eq:complex} for $\rho=\rho_2'$) is estimated as the time to reach the abundance $x_e+\Delta x$ for small $\Delta x$:
\begin{equation}\label{eq:collapse}
T_{col}=\tau(x_e+\Delta x;\vv{p})-\tau(x_2;\vv{p}),
\end{equation}
where $x_e$, $\gamma$ and $\beta$ are computed at $\vv{p}=(\rho_2+\Delta\rho,\mu,\sigma)$. 
\item[(ii)]  \emph{Recovery time}. Now the initial condition is $x=x_1$. Catches are slightly reduced from $\rho_1$ to $\rho_1'=\rho_1-\Delta\rho$ 
with $\Delta\rho$ close to zero. We then estimate the recovery time as the time to reach the abundance $x_e-\Delta x$ for small $\Delta x$, $x_e$
being the only real root of Eq.~\eqref{eq:complex} for $\rho=\rho_1'$:
\begin{equation}\label{eq:recovery}
T_{rec}=\tau(x_e-\Delta x;\vv{p})-\tau(x_1;\vv{p}),
\end{equation}
where $x_e$, $\gamma$ and $\beta$ are calculated at $\vv{p}=(\rho_1-\Delta\rho,\mu,\sigma)$. 
\end{itemize}

We have to use the approximated endpoints $x_e+\Delta x$ for collapse and $x_e-\Delta x$ for recovery times
instead of the actual equilibrium point $x_e$ because the time to reach $x_e$ is infinite. Calculation of $T_{col}$ and $T_{rec}$ only 
amounts to solve numerically the cubic equation~\eqref{eq:complex} for $x_e$, $\gamma$ and $\beta$.

Results for collapse and recovery times are summarized in 
Fig.~\ref{fig:deterministic} (right panel) as a function of immigration. We observe two remarkable phenomena: (i) for immigration large enough, the time 
to collapse is always larger than the time to recover ---this occurs because collapse times are measured at values of the harvesting rate 
($\rho\approx \rho_2$) larger than those of recovery ($\rho\approx \rho_1<\rho_2$) and times depend on these parameters in a non-trivial way. 
(ii) Although individuals are supplied to the system at higher rates for increasing immigration, it takes more time to collapse and (surprisingly) to recover 
the system. As before, this effect is caused by the monotonic increase in $\rho_{1,2}$ as $\mu$ augments (Fig.~\ref{fig:deterministic}, middle panel). This is 
a fingerprint of being close to a critical transition: as immigration gets closer to the critical value $\mu_c$, the dynamics of the system becomes stiffer and 
takes more time from going from the initial condition to the endpoint~\citep{scheffer:2012}. Collapse and recovery times therefore behave as early 
warning signals of the transition between a fully reversible system ($\mu>\mu_c$) and a population undergoing collapses and recoveries in population 
abundances ($\mu<\mu_c$).

It is possible to find global, analytical lower bounds for collapse and recovery times. We leave the derivation for Appendix~\ref{sec:appB} and report the
results here: we can approximate the collapse time as
\begin{equation}\label{eq:colapp}
T_{col}\approx \frac{\pi\sigma\left[\sigma+5+2(\sigma-1)\cos\left(\frac{1}{3}\cos^{-1}\left(1-\frac{2\mu}{\mu_c}\right)\right)\right]}
{2(\sigma-1)\sqrt{\sigma\Delta\rho\left[2\cos\left(\frac{2}{3}\cos^{-1}\left(1-\frac{2\mu}{\mu_c}\right)\right)+1\right]}},
\end{equation} 
and the recovery time as
\begin{equation}\label{eq:recapp}
T_{rec}\approx \frac{\pi\sigma\left[\sigma+5-2(\sigma-1)\sin\left(\frac{1}{3}\sin^{-1}\left(1-\frac{2\mu}{\mu_c}\right)\right)\right]}
{2(\sigma-1)\sqrt{\sigma\Delta\rho\left[2\cos\left(\frac{2}{3}\sin^{-1}\left(1-\frac{2\mu}{\mu_c}\right)\right)-1\right]}}.
\end{equation}
Figure~\ref{fig:deterministic} (right panel) shows that these approximations are indeed accurate lower bounds for actual times. The 
estimate~\eqref{eq:colapp} is especially good for collapse times at small values of immigration. 

Several insightful limits can be derived from these formulae: (i) $T_{col}$ and $T_{rec}$ diverge as $\mu\to\mu_c^{-}$, and they grow faster than 
\begin{equation}
T_c:=\frac{\pi}{2}\sqrt{\frac{\sigma}{\Delta\rho}}\left(\frac{\sigma+2}{\sigma-1}\right)\left[3\left(1-\frac{\mu}{\mu_c}\right)\right]^{-1/4},
\end{equation}
where $T_c$ has been obtained by a power series approximation of (\ref{eq:recapp}) and (\ref{eq:colapp})  about $\mu_c$, both coinciding.
This divergence suggests that collapse and recovery times can be regarded as early warning signals for the critical transition at $\mu=\mu_c$.
(ii) In the small immigration regime ($\mu\ll\mu_c$), collapse times remain finite,
\begin{equation}\label{eq:colappmusmall}
T_{col}\approx \frac{\pi}{2}\sqrt{\frac{\sigma}{\Delta\rho}}\left(\frac{\sigma+1}{\sigma-1}\right)\left[1+\frac{4}{27}
\left(\frac{\sigma+3}{\sigma+1}\right)\frac{\mu}{\mu_c}\right],
\end{equation}
whereas recovery times diverge as $\mu$ goes to zero,
\begin{equation}\label{eq:recpmusmall}
T_{rec}\approx \frac{\pi}{2(\sigma-1)}\sqrt{\frac{\sigma}{\Delta\rho}}\left(\frac{27\mu_c}{\mu}\right)^{1/4}.
\end{equation}
Recovery times diverge as $\mu^{-1/4}$ both for $\mu\to 0^+$ and $\mu\to\mu_c^-$. 
%(iii) From (\ref{eq:colappmusmall}) and (\ref{eq:recpmusmall}) it can be seen that $T_{col}$ increases with $\mu$ while $T_{rec}$  decreases, being this monotonicity valid  as far as the immigration rate is kept below the threshold $\mu_k:=\frac{(-1+\sigma)^3(3+\sigma)}{\sigma(5+\sigma)^3}$ ($\mu <\mu_k$). On the contrary of the surprising  behaviour of $T_{rec}$ when $\mu \in [\mu_k,\mu_c]$, its  decrease  in the small immigration regime adjust better to a predictable scenery.
And (iii) as expected, both times diverge in the limit $\Delta\rho\to 0$. The functional form of this divergence is precisely $(\Delta\rho)^{-1/2}$. 

The latter observation can be used to define \emph{scaled collapse and recovery times} that do not explicitly depend on the separation $\Delta\rho$. 
In fact, the limits $\displaystyle\widetilde{T}_{col}:=\lim_{\Delta\rho\to 0^+} \sqrt{\Delta\rho}\,T_{col}$ and 
$\displaystyle\widetilde{T}_{rec}:=\lim_{\Delta\rho\to 0^+} \sqrt{\Delta\rho}\,T_{rec}$ are finite, independent of $\Delta\rho$ 
and $\Delta x$. The only dependence that remains in these scaled times is through model parameters $\mu$ and $\sigma$.

\section{Stochastic collapse and recovery times}
\label{sec:stochastic}

Populations in nature are discrete. Growth and death events, as well as interactions, proceed by discrete variations in the number of individuals.
The effect introduced by demographic stochasticity in the population dynamics can make discrete abundances depart substantially from to the 
values predicted by deterministic models~\citep{capitan:2015,capitan:2017}. Our main goal is to estimate collapse and recovery times in the presence 
of demographic stochasticity, and compare them to their deterministic counterparts.

\subsection{Demographic stochasticity}
\label{ssec:demstoch}

A routinely-used methodology to formulate stochastic models whose deterministic limit is a given differential equation is based on identifying 
deterministic growth and death rates and set them as the transition probability rates of a stochastic, continuous-time Markov 
process~\citep{haegeman:2011}. Let $n$ be the population abundance, i.e., the number of individuals in the system ($n=0,1,2,\dots$). Elementary 
processes that increase the number of individuals are intrinsic births and immigration, whereas the process that lower population abundance are 
intrinsic deaths, competition among individuals and harvesting. The suitable mathematical formulation for a population subject to stochastic increases 
and decreases of the number of individuals is precisely a birth-death process.

Birth-death processes are completely defined once we specify explicit forms for the probabilities per unit time $q^+_n$ and $q^-_n$ that the 
transitions $n\to n+1$ (`births') and $n\to n-1$ (`deaths') occur, respectively. To reproduce the dynamics~\eqref{eq:immi} we choose 
\begin{equation}
q^+_n=r^+n+I,
\end{equation}
because intrinsic births are proportional to abundances, contrary to immigration events, which occur at constant rate. Here $r^+n$ is the probability rate that 
an intrinsic birth takes place. Similarly, we set the death rate as
\begin{equation}
q^-_n=r^-n+\frac{rn^2}{K}+\frac{cn}{1+bn}
\end{equation}
because intrinsic deaths are proportional to abundances, competition among individuals is a quadratic term and harvesting includes saturation. 
In order to reproduce model~\eqref{eq:immi} in the deterministic limit we have to set $r=r^+-r^-$~\citep{haegeman:2011,capitan:2015}. Here we assume 
$r^+>r^-$ so that the balance of rates $r=r^+-r^-$ is positive and yields growth in the absence of competition and harvesting.

With these rates, the probability $p_n(t)$ of observing $n$ individuals at time $t$ is described by an infinite system of coupled differential equations (the
so-called forward-Kolmogorov master equation),
\begin{equation}\label{eq:master}
\frac{d p_n(t)}{d t}=q^+_{n-1}p_{n-1}(t)+q^-_{n+1}p_{n+1}(t)-(q^+_n+q^-_n)p_n(t),
\end{equation}
which simply establishes the balance of probability due to birth and death events. Thanks to a~\cite{vankampen:1992} series expansion
of the master equation in the system size, it can be shown that the deterministic limit of the stochastic process is precisely Eq.~\eqref{eq:immi}.
Although for small values of $K$ both dynamics can greatly differ, it holds that the stochastic dynamics converges to the deterministic one in the
limit $K\to\infty$ (we prove this statement in Appendix~\ref{sec:appC} using mean first-passage times, which are defined in a subsection below). In what
follows we assume that the net effect of demographic stochasticity is encoded in $K$ (smaller carrying capacities lead to systems with larger 
demographic noise).

The master equation can be re-scaled as we did for the deterministic equation. We scale integer abundances as $z:=bn$ (we call this new variable
$z$ as $z$-abundance). Multiplying both sides of Eq.~\eqref{eq:master} by $b/r$ and defining $\tau:=rt$ and $P(z,\tau):=p_{z/b}(\tau/r)$, we obtain 
the PDE
\begin{align}\label{eq:master2}
b\frac{\partial P(z,\tau)}{\partial \tau}=&q^+(z-b)P(z-b,\tau)\nonumber\\
&+q^-(z+b)P(z+b,\tau)\nonumber\\
&-[q^+(z)+q^-(z)]P(z,\tau),
\end{align}
where we have set
\begin{equation}\label{eq:qpm} 
\begin{aligned}
&q^+(z) := \frac{b\,q^+_n}{r} = \alpha z+\mu,\\
&q^-(z) := \frac{b\,q^-_n}{r} = (\alpha -1)z+\dfrac{z^2}{\sigma}+\dfrac{\rho z}{1+z},
\end{aligned}
\end{equation}
$\alpha$ has been defined as the ratio $\alpha:=r^+/r\geq 1$, and we have used that $r^-/r=\alpha-1$. As for the deterministic model, scaled 
parameters are here defined as $\sigma=Kb$, $\rho=c/r$ and $\mu=bI/r$.

Below we investigate the existence of a bifurcation in the stochastic scenario, and provide a methodology to measure collapse and recovery 
times in the presence of demographic stochasticity. 

\subsection{Steady-state probability distribution}
\label{ssec:stochbif}

\begin{figure*}[t!]
\begin{center}
\hspace*{-5mm}
\includegraphics[width=1.03\textwidth]{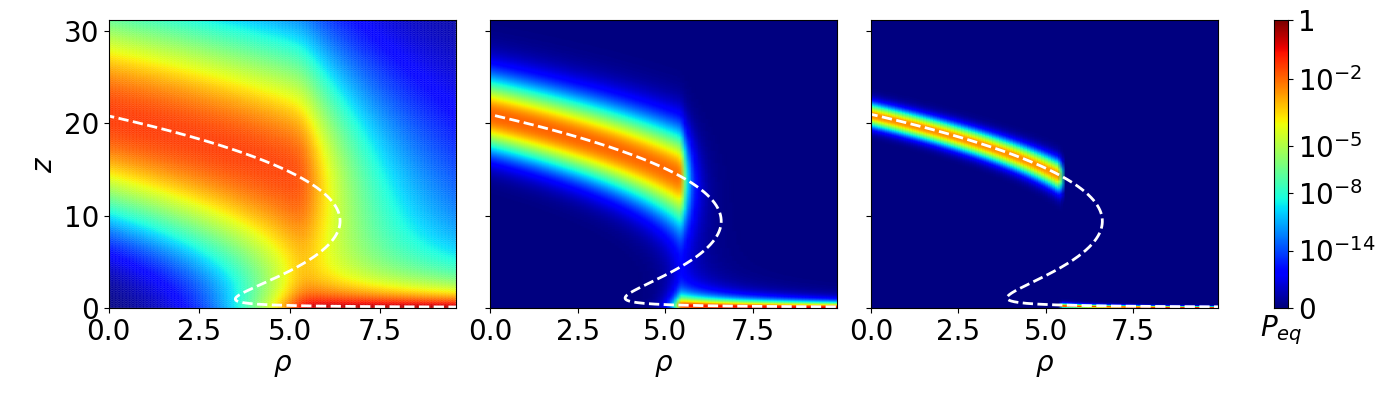}
\caption{
\label{fig:probs} 
\emph{Steady-state distribution heatmap} for increasing catching efforts $\rho$ and three carrying capacity values, $K=100$ (left), $K=1000$ (middle) and 
$K=10000$ (right). Probability values $P(z)$ are encoded in the colorbar, and represented as a function of $z$ in the vertical axis. To ease visualization
by highlighting small numbers, probabilities have been transformed to $\widetilde{P}(z)=[P(z)]^{1/20}$ ---the scale of the colorbar shows
actual probability values, though. Note that the two maxima are very well predicted by Eq.~\eqref{eq:stocubic} outside the bi-stability range. As $K$ 
augments, distributions get more and more peaked around the solutions of Eq.~\eqref{eq:stocubic}, which is shown as a white, dashed line. Remaining 
parameter values are $\sigma=20$, $\mu=1$ and $\alpha=1$.
}
\end{center}
\end{figure*}

First we study some equilibrium properties of the stochastic model to determine whether the bifurcation previously described maintains or not. The 
steady-state probability distribution (also known as equilibrium distribution) is defined as $\displaystyle P(z):=\lim_{\substack{\tau\to\infty}} P(z,\tau)$. 
If this limit exists, the distribution is independent of $\tau$ and therefore the rhs of the master equation~\eqref{eq:master2} has to be equal to zero. 
This implies the detailed balance condition,
\begin{align}\label{eq:balance}
q^-(z&+b)P(z+b)-q^+(z)P(z)=\nonumber \\
&=q^-(z)P(z)-q^+(z-b)P(z-b),
\end{align}
which yields the following recurrence relation to be satisfied by the equilibrium distribution,
\begin{equation}\label{eq:rec}
P(z)=\frac{q^+(z-b)}{q^-(z)}P(z-b).
\end{equation}
Eq.~\eqref{eq:rec} determines (numerically) the steady-state probability distribution up to a multiplicative factor which can be calculated by normalization. 
Equilibrium distributions as a function of the harvesting rate $\rho$ are depicted in Fig.~\ref{fig:probs} for increasing values of the carrying capacity $K$. 

Depending on model parameters, equilibrium distributions may exhibit a single maximum or two local maxima and a local minimum in between. In order to 
find these extrema, as in~\cite{capitan:2017} we expand $P(z-b)$ about $z$ up to first order and look for solutions of~\eqref{eq:rec} that verify $P'(z)=0$: 
imposing $P(z-b)=P(z)+O(b)^2$ in Eq.~\eqref{eq:rec} yields $q^+(z-b)\approx q^-(z)$, which is equivalent to
\begin{equation}\label{eq:stocubic}
z\left(1-\frac{z}{\sigma}\right)-\frac{\rho z}{1+z}+\mu-\alpha b=0.
\end{equation}
This condition is formally equal to its deterministic counterpart, Eq.~\eqref{eq:cubic}, with an \emph{effective immigration rate}
$\mu':=\mu-\alpha b=\mu-\frac{\alpha\sigma}{K}$ instead. Accordingly, in terms of rho, the new cubic curve yields ranges with a single maximum in P(z) (which would correspond to the upper and lower single equilibrium states in the deterministic model) and ranges with two maxima and a minimum between them (corresponding to the two stable and the unstable equilibria of the deterministic model, respectively)\footnote{Observe that effective immigration rate can be negative or zero. In those cases, $P(z)$ has a local extremum in the boundary $z=0$, corresponding to a maximum when $P(0)>P(b)$ [or, equivalently, when $q^+(0) < q^-(b)$] and a minimum otherwise. Let $\rho^{\star}=(1+b)\left(1-\frac{b}{\sigma}+\frac{\mu-\alpha b}{b}\right)$ be the harvesting rate satisfying $q^+(0)=q^-(b)$, then when $\mu'\le 0$ $P(z)$ has a local maximum at $z=0$ for $\rho>\max(\rho^{\star},0)$ and a minimum otherwise.}.
Effective immigration takes into account stochasticity through parameter $K$ for fixed scaled carrying capacity 
$\sigma$. As we recover the deterministic limit for $K\to\infty$, the lower the value of $K$ the larger the differences between the equilibrium points 
of the deterministic dynamics and the critical points of $P(z)$ in the presence of stochasticity. Consistently, as $K$ increases, distributions get closer to 
the curve given by Eq.~\eqref{eq:stocubic}, see Fig.~\ref{fig:probs}.

According to steady-state probabilities, it seems that Eq.~\eqref{eq:stocubic} does not capture well the range at which two maxima coexist
(Fig.~\ref{fig:probs}). Equilibrium probabilities fail to capture the locations of the two maxima at a time: probabilities tend to accumulate just around one of 
them. Therefore, the steady-state distribution does not match perfectly the predicted cubic curve, and based \emph{solely} on that distribution we do not 
observe a clear-cut bifurcation ---contrary to what happened in the deterministic case. In the following subsection we show that, when measuring collapse 
and recovery times in terms of the time taken by the system to escape from one local maxima and reach the other, a bifurcation-like framework re-enters 
the scene.

\subsection{Stochastic collapse and recovery times}
\label{ssec:stochtimes}

Our definition of collapse and recovery times in the presence of demographic stochasticity relies on mean first-passage times. The \emph{first-passage 
time} of a Markov process $n(t)$ is the random variable $t$ defined as the time that the process takes to first hit state $n_1$ given that the 
process was initiated in state $n(0)=n_0$. Let $f(t|n_0\to n_1)$ be the pdf of the random variable $t$, hence the mean first-passage time is defined
as
\begin{equation}
\bar{t}_{n_0,n_1}:=\int_0^{\infty} t f(t|n_0\to n_1) dt.
\end{equation}
Following~\cite{gillespie:1991}, and using the backward-Kolmogorov master equation formalism, mean first-passage times can be calculated in terms of 
probability rates $q^+_n$ and $q^-_n$ and the steady-state probability distribution $p_n$ as
\begin{equation}\label{eq:fpt}
\bar{t}_{n_0,n_1} = \begin{cases}
\displaystyle\sum_{\ell=n_0}^{n_1-1}\frac{\sum_{j=0}^{\ell}p_j}{q^+_{\ell} p_{\ell}}  & \text{if\  } n_0 < n_1,\\
\displaystyle\sum_{\ell=n_1}^{n_0-1}\frac{1-\sum_{j=0}^{\ell}p_j}{q^+_{\ell}p_{\ell}} & \text{if\  } n_1 < n_0.
\end{cases}
\end{equation} 
Eq.~\eqref{eq:fpt} can be derived by showing that the backward-Kolmogorov equation can be written as~\citep{gillespie:1991}
\begin{equation}\label{eq:fptrec}
q_{n_0}^+(\bar{t}_{n_0,n_1}-\bar{t}_{n_0+1,n_1})+q_{n_0}^-(\bar{t}_{n_0,n_1}-\bar{t}_{n_0-1,n_1})=1
\end{equation}
and then get~\eqref{eq:fpt} by iteration of the above recurrence relation. Although~\eqref{eq:fpt} involves the steady-state distribution to compute
$\bar{t}_{n_0,n_1}$, the expression is valid  for the mean time to hit $n_1$ starting from $n_0$, irrespective of the process being at stationarity
or not.

Now we use Eq.~\eqref{eq:stocubic} to define stochastic collapse and recovery times. For a fixed immigration rate, let $\rho_{1,2}$ be the lower and
upper limits of the range where two maxima coexist in the steady-state probability distribution, and $z_{1,2}$ the scaled abundances at these limits 
(observe that then Eqs.~\eqref{eq:x12} and~\eqref{eq:rho12} hold evaluated at the effective immigration $\mu'=\mu-\alpha b$ instead of $\mu$). 
Both $z_{1,2}$ correspond to inflection points of the distribution at the edges $\rho_{1,2}$. At $\rho_{1,2}$ the 
abundance of the local maximum $z_e$ can also be calculated analytically (cf. Eqs.~\eqref{eq:x3r} and~\eqref{eq:x3c} in Appendix~\ref{sec:appB} 
with $\mu'$ instead of $\mu$). Going back to original integer abundances, associated to each value of $z$ we define the corresponding integer abundance 
as $n=\lfloor\frac{z}{b}\rfloor$. Let us denote the $n$-abundance at the lower-branch critical point as $n_1=\lfloor\frac{z_1}{b}\rfloor$, and 
$n_2=\lfloor\frac{z_2}{b}\rfloor$ the $n$-abundance at the upper branch. Then we define the \emph{stochastic collapse time} as the mean 
first-passage time $T_{col}=r\bar{t}_{n_2,n_e'}=:\bar{\tau}(z_2,z_e+\Delta z)$ when catching effort is increased from $\rho_2$ to $\rho_2+\Delta\rho$, 
and the final state is $n_e'=\lfloor\frac{z_e+\Delta z}{b}\rfloor$ with $\Delta z$ a small abundance (observe that we used scaled times $\tau=rt$ to 
make them comparable with deterministic ones). This definition is a literal transcription, to the language of the stochastic model, of the definition we 
adopted in subsection~\ref{ssec:times}. Similarly, the \emph{stochastic recovery time} is defined as the mean first-passage time 
$T_{rec}=r\bar{t}_{n_1,n_e'} =:\bar{\tau}(z_1,z_e-\Delta z)$ when catching effort is decreased from $\rho_1$ to $\rho_1-\Delta\rho$, and the endstate is 
$n_e'=\lfloor\frac{z_e-\Delta z}{b}\rfloor$.

\begin{figure*}[t!]
\begin{center}
\hspace*{-5mm}
\includegraphics[width=1.03\textwidth]{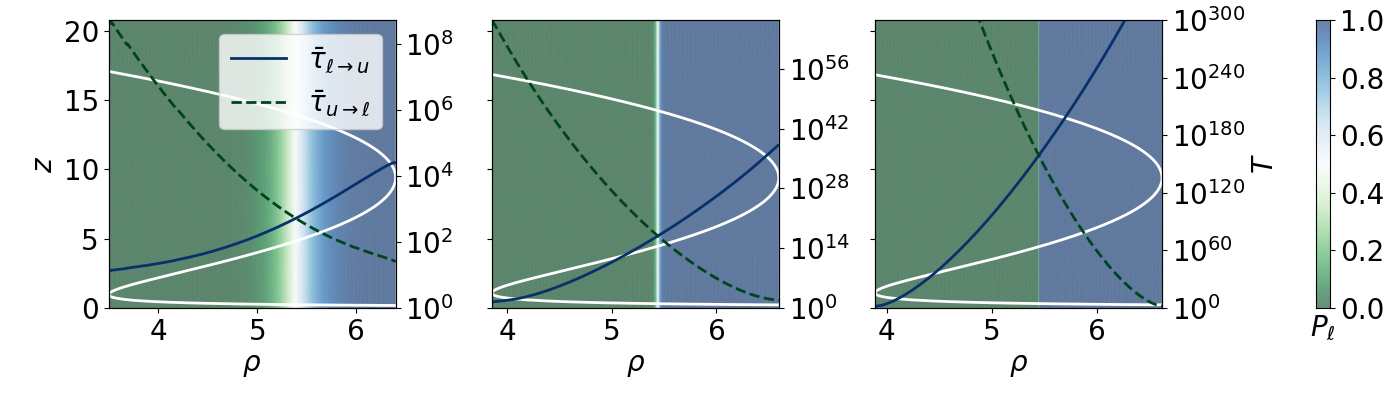}
\caption{
\label{fig:freqs} 
\emph{Mean first-passage times calculated for collapse and recovery transitions} are depicted for decreasing stochasticity ($K=100$, left panel; 
$K=1000$, middle panel; $K=10000$, right panel) as functions of catching effort. The heatmap represents the aggregated probability $P_{\ell}:=\sum_{n\le 
n_{m}} p_n$, $p_n$ being the steady-state distribution and $n_{m}$ the $n$-abundance associated to the intermediate \emph{minima} of that distribution. 
We interpret $P_{\ell}$ as an aggregated stationary probability for the system to remain close to the collapsed state ($n\approx 0$). $P_{\ell}$ exhibits a 
sharp transition at an intermediate harvesting rate, which becomes sharper as $K$ increases. However, even in the blue area, where the probability 
$P_{\ell}$ of observing a collapsed state is almost one in the limit $t\to\infty$, it takes a very long time for the system to collapse starting from the upper 
sbranch of maxima. Similarly, the time to escape the lower maxima in the green region (where the probability of observing the system near the upper
maximum is close to one) is very high. In all panels, solutions of Eq.~\eqref{eq:stocubic} are marked with white color, and we take $\sigma=20$, $\mu=1$ 
and $\alpha=1$.
}
\end{center}
\end{figure*}

Are these definitions sound, especially taking into account that, \emph{in the limit $t\to\infty$}, most of the probability is concentrated in one of the two
maxima? Figure~\ref{fig:freqs} shows that this is the case, in spite of the picture portrayed by the steady-state distribution. Observe first that we can 
calculate mean first-passage times for collapse and recovery transitions for any intermediate value $\rho_1 < \rho < \rho_2$ of the harvesting rate, 
because the Markov process is ergodic and all states are visited with non-zero probability. In those cases, we simply solve numerically the cubic 
equation~\eqref{eq:stocubic} to compute $n$-abundances for the two maxima (upper and lower abundances $n_u$ and $n_{\ell}$, respectively), and 
calculate the mean first-passage times $\bar{\tau}_{u\to \ell}$ and $\bar{\tau}_{\ell\to u}$ for the transitions $n_u\to n_{\ell}$ (collapse) and 
$n_{\ell}\to n_u$ (recovery). These curves are plotted in Fig.~\ref{fig:freqs} as functions of $\rho$. We observe that $\bar{\tau}_{u\to \ell}$ 
($\bar{\tau}_{\ell\to u}$) grows faster than exponentially as $\rho$ decreases (increases), and takes astronomically large values even for moderate 
values of the carrying capacity $K$. This phenomenon occurs for almost any value of the catching effort within the range $(\rho_1,\rho_2)$. This 
means that the time to collapse from the upper branch is very large and remains large until the edge $\rho_2$. Although the probability of visiting the 
upper maxima is close to zero \emph{in the limit} $t\to\infty$ when $\rho\lesssim\rho_2$, the system takes a very long time to collapse to the lower 
maximum, where most of the probability is actually accumulated at stationarity. Thus, the process remains at a \emph{quasi-stationary} state in the upper 
branch as we increase the catching effort, and this quasi-stationary state extends much further in $\rho$ than expected according to equilibrium 
probabilities (Fig.~\ref{fig:probs}). Once $\rho$ is very close to $\rho_2$, collapse times are small and the system effectively decays to the lower-branch 
maxima. The exact same picture is observed for the reverse transition in a recovery scenario, when $\rho$ decreases along the lower branch of maxima. 
Therefore, the stochastic process exhibits a bifurcation similar to that described for the deterministic model, and Eq.~\eqref{eq:stocubic} can be used to 
determine initial and final states before collapse and recovery.

\begin{figure*}[t!]
\begin{center}
\hspace*{-5mm}
\includegraphics[width=1.03\textwidth]{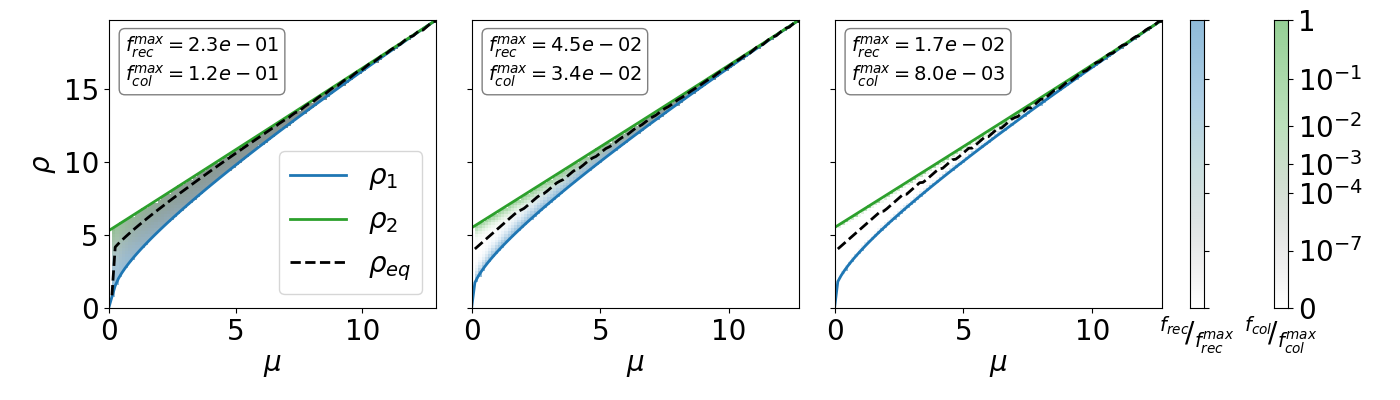}
\caption{
\label{fig:phase} 
\emph{Collapse and recovery frequencies.} The stochastic predictions for the limits of the bi-stability region are depicted in the $(\mu,\rho)$
plane for $\sigma=20$, $\alpha=1$ and $K=100$ (left), $K=1000$ (middle), $K=10000$ (right). Heatmaps stand for collapse and recovery
frequencies, which for the sake of comparison have been normalized dividing by the maximum values they take along the interval $\mu\in[0,\mu_c]$,
which are specified in legends. The dashed line represents the curve $\rho=\rho_{eq}(\mu)$ for which $P_{\ell}=\frac{1}{2}$ ---note that this point
coincides remarkably well with the value of $\rho$ at which $\bar{\tau}_{u\to\ell}$ equals $\bar{\tau}_{\ell\to u}$, see Fig.~\ref{fig:freqs}. As $K$ 
increases, collapse and recovery frequencies tend to concentrate along the upper or lower branches $\rho_1(\mu)$ and $\rho_2(\mu)$, respectively.
}
\end{center}
\end{figure*}

We can check quantitatively that the system indeed remains in a quasi-stationary state along the curves $\rho_{1,2}(\mu)$ that determine the location of
the two maxima, especially for large carrying capacities ---natural populations usually operate at large $K$ values, in particular, 
large fisheries~\citep{maroto:2014}. For this purpose we screen the parameter space by varying $\rho$ within the range $\rho_1(\mu)\le \rho\le \rho_2(\mu)
$ for $0\le \mu\le \mu_c$ (observe that, in the stochastic model, the critical immigration rate is $\mu_c=\frac{1}{\sigma}\left(\frac{\sigma-1}{3}\right)^3+\alpha 
b$) and calculate collapse and recovery times. Figure~\ref{fig:phase} reports results for the average first-passage frequencies for collapse and recovery 
transitions, defined as $f_{col}:=\bar{\tau}_{u\to\ell}^{-1}$ and $f_{rec}=\bar{\tau}_{\ell\to u}^{-1}$, respectively. It is clear from Fig.~\ref{fig:phase} that, as 
long as carrying capacity increases, collapse and recovery transitions between the upper and lower branches become more and more unfrequent. This 
confirms quantitatively that the system remains at a quasi-stationary state, the cubic equation~\eqref{eq:stocubic} remains valid to describe system 
transitions along a hysteresis cycle, and the mathematical description of the bifurcation is very similar to the deterministic one analyzed in 
Section~\ref{sec:deterministic}.

\begin{figure*}[t!]
\begin{center}
\hspace*{-5mm}
\includegraphics[width=1.03\textwidth]{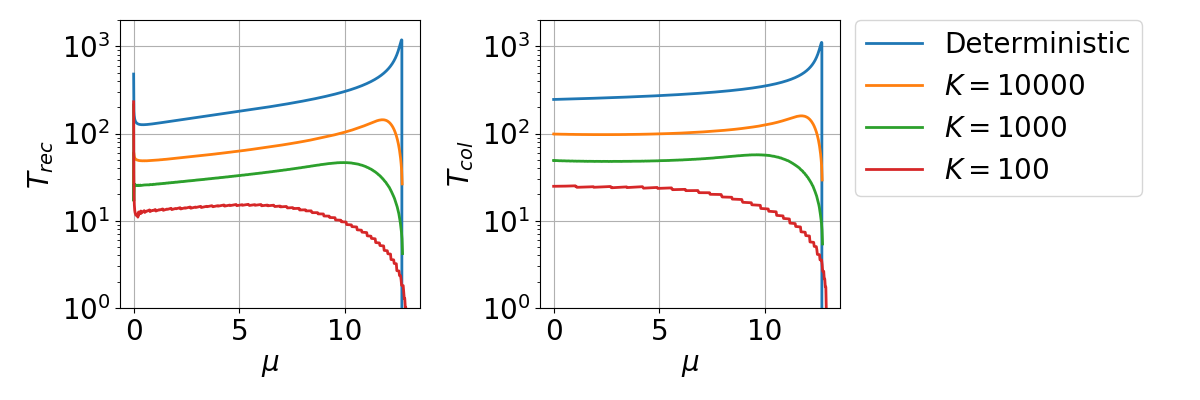}
\caption{
\label{fig:times} 
\emph{Stochastic collapse and recovery times as a function of immigration.} Stochastic times are upper-bounded by the deterministic ones for almost
all the range of immigration rates, and tend to approximate in values and even in shape to deterministic times as carrying capacity increases. 
Remaining model parameters are $\sigma=20$, $\alpha=1$, $\Delta\rho=0.001$ and $\Delta x=0.01$.
}
\end{center}
\end{figure*}

So far we have proved that the deterministic definition of collapse and recovery times can be soundly extended to situations where demographic 
stochasticity is present. In order to consistently compare stochastic times with their deterministic counterparts, in Appendix~\ref{sec:appC}
we show that, by taking the limit of large carrying capacity in Eq.~\eqref{eq:fptrec}, we recover the deterministic dynamics given by Eq.~\eqref{eq:imscal}.
This comparison is presented in Fig.~\ref{fig:times}. As a function of the carrying capacity, both collapse and recovery times increase with $K$ globally
for all the range in immigration. 

\section{Discussion}
\label{sec:disc}

Characterizing temporal dynamics in ecological systems that undergo shifts and transitions is crucial for their control and management, especially
if those systems are subject to exploitation. Inspired by population models traditionally studied in fisheries~\citep{clark:1979}, we provided estimates 
for the times that a harvested population takes to collapse and to recover when catching efforts are varied along a cycle. Compared to previous work, here
we thoroughly studied the effect of an external source of individuals (immigration), and characterized the pitchfork-like bifurcation of the dynamical
system in two different contexts: (i) when populations are discrete and demographic stochasticity plays a role, and (ii) in the (deterministic) limit of large
carrying capacities. In the latter we were able to provide analytically precise, consistent lower bounds for collapse and recovery times, and in the
presence of demographic stochasticity we unveiled a quasi-stationary state that justified using a similar cubic equation for initial and final states in that 
setting. Finally, we numerically quantified the variation of stochastic collapse and recovery times with immigration.

An important implication of our work is that we can use the two maxima of the steady-state probability distribution as initial states to collapse and 
recovery transitions in the presence of stochasticity. We interpreted the astronomically large time associated to collapse transitions starting from the 
upper maximum $n_u$ as a quasi-stationary state that permitted us to measure collapse times as the mean first-passage time associated to the 
transition from $n_u$ to $n_{\ell}$. One can imagine this stochastic system as described by two coarse-grained states $u$ and $\ell$, associated to the 
two maxima of the distribution, and pose the dynamics of the system as a two-state continuous-time Markov process defined by the probability rates 
proportional to frequencies associated to collapse and recovery mean first-passage times, $q_{u\to\ell}=\bar{\tau}_{u\to\ell}^{-1}$ and 
$q_{\ell\to u}=\bar{\tau}_{\ell\to u}^{-1}$. The steady-state distribution of this coarse-grained system is trivially calculated as 
\begin{equation}
\begin{aligned}
&p_{u}=\frac{q_{u\to\ell}}{q_{u\to\ell}+q_{\ell\to u}}=\frac{f_{col}}{f_{col}+f_{rec}},\\
&p_{\ell}=\frac{q_{\ell\to u}}{q_{u\to\ell}+q_{\ell\to u}}=\frac{f_{rec}}{f_{col}+f_{rec}}.
\end{aligned}
\end{equation} 
Then it can be checked that these probabilities almost perfectly overlap the probabilities $P_{u}$ and $P_{\ell}$ defined by aggregating the
steady-state distribution $p_n$ of the original Markov process, $P_{\ell}:=\sum_{n\le n_{m}} p_n$ and $P_{u}:=\sum_{n> n_{m}} p_n$, $n_m$
being the abundance of the local minimum of the equilibrium distribution (results not shown). This fact reinforces our result that the system behaves 
as if it was formed by two bi-stable, coarse-grained states, which allows to define collapse and recovery times as in the deterministic scenario. 

In the presence of demographic stochasticity, which leads to increased extinctions through ecological drift when compared to deterministic 
scenarios~\citep{capitan:2015}, stochastic collapse and recovery times were found smaller than deterministic times for the whole range in immigration
rates. This counterintuitive effect can be interpreted as follows: for small $K$, actual discrete $n$-abundances for the two maxima $n_u$ and $n_{\ell}$ 
are small because average system's size is controlled by $K$, and the larger the carrying capacity the larger abundances can be in practice. Therefore, 
for small $K$ the stochastic process needs on average less discrete steps (mediated by birth or death events) to go from $n_u$ to $n_{\ell}$ or 
\emph{vice versa}, and consequently it takes less time to collapse or recover. The fact that stochastic times are always upper bounded by deterministic 
ones can be regarded as a \emph{finite size effect} attributable to finite carrying capacities. In Appendix~\ref{sec:appC} we prove that mean first-passage 
times derived from Eq.~\eqref{eq:fptrec} actually converge to the deterministic dynamics~\eqref{eq:imscal} as the carrying capacity grows to infinity, 
showing that large-sized system are progressively less affected by stochasticity and the system is better approximated with the deterministic model. 
Therefore, as well as a measure of system size, finite values of $K$ control the magnitude of demographic stochasticity in the process.

The fact that stochastic times are smaller than their deterministic counterparts has important practical implications in the maintenance of harvested
ecosystems. In systems with finite size, and for the same levels of external immigration, it takes less time to undergo a collapse transition than the
expected time according to the corresponding deterministic dynamics. This is important because state-or-the-art models used in the context of
ecosystems subject to exploitation tend to use routinely predictions based on deterministic dynamics which, as we prove, lead to overestimation in the
time to reach collapsed equilibria close to extinction. On the positive side, stochastic recovery times are smaller than deterministic ones as well. 
Therefore it takes less time to recover the original population level once the system has trespassed the threshold $\rho_1$ at which recovery
takes place.

At first glance, it is surprising that collapse and recovery times both increase with immigration. Individuals are being supplied to the system
at increasing rates so that, in principle, smaller recovery times would be expected. However, collapse and recovery times depend also implicitly 
on the thresholds $\rho_{1,2}$ where collapse and recovery transitions take place, and these limits in catching efforts both grow with immigration. 
This means that the negative effect of increased harvesting rates balances the positive effect of adding new individuals and dominates over it. 
In addition, we expect that the dynamics becomes stiffer as immigration rates approach $\mu_c$, as routinely occurs in systems approaching 
critical transitions [it is the so-called ``critical slowing down'' effect near tipping points, see~\cite{scheffer:2012} and references therein], so we expect 
growing collapse and recovery times as immigration increase, and our estimates and analytical results for collapse and recovery times could be used in
practice to anticipate the critical transition above the tipping point in immigration. It is our expectation that stochastic collapse and recovery times 
would increase and decrease, respectively, with immigration if we measure them using a fixed harvesting rate $\rho$. We leave this study for future work.

Our estimates for the collapse and recovery times include the combined effect of augmenting catching effort and immigration rate: as new individuals are 
supplied to the population when $\mu$ increases, the harvesting rates $\rho_{1,2}$ increase as well but the net effect is that collapse and recovery times 
are larger (Fig.~\ref{fig:times}). This has important implications for fisheries management: repopulating a certain area with a native species can help to 
prevent collapses in the sense that the time taken to collapse increases as $\mu$ increase. However, the time to recovery is also larger for augmented 
immigration: therefore, after a collapse, repopulating the fishery can have a opposing effect to population recovery. Although the effect of adding new 
individuals to fisheries in real situations is unclear even under moratorium restrictions~\citep{loria:1991}, our results suggest that repopulating areas at 
smaller rates may help to accelerate the recovery of the system. There is, however, a threshold in immigration below which recovery times start increasing: 
as we showed analytically and numerically, recovery times reach a minimum value both in deterministic and stochastic scenarios (see 
Fig.~\ref{fig:deterministic}, right panel, and Fig.~\ref{fig:times}). Let $\mu_m$ be the immigration rate at which $T_{rec}$ reaches its minimum value. 
We thus expect that immigration rate has to be fine-tuned to be close to $\mu_m$ for a collapsed system to recover as fast as possible. Recovery 
strategies based on moratoriums [see for example~\cite{rose:2015} for cod] operate on a natural (low immigration) dispersal regime, which (under our 
modeling assumptions) may lead to diverging recovery times. A research program of capital (and very practical) importance for fishery management would 
be to investigate whether actual repopulation policies are optimal according to our predictions.

\section{Acknowledgements}
\label{sec:acknow}
This contribution was conceived during a fruitful discussion that came about at Gema Lucero's place after an aikido practice. JAC acknowledges financial 
support from Ministerio de Econom{\'\i}a y Competitividad projects BRIDGES (CGL2015-69043-P) and CRISIS (PGC2018-096577-B-I00).

\section*{Appendices}

\renewcommand\thesection{\Alph{section}}
\setcounter{section}{0}
\renewcommand\theequation{\Alph{section}.\arabic{equation}}

\section{Analytical expressions for $x_{1,2}$}
\label{sec:appA}
\setcounter{equation}{0}

In this appendix we find analytical expressions for the non-negative solutions $x_{1,2}$ of Eq.~\eqref{eq:rhop} using Cardano's method for solving
cubic equations. First define $z:=x/(\sigma-1)$ and $\nu:=\mu/\mu_c$ as re-scaled abundance and immigration ---this way, $\nu$ is standardized to 
take values between $0$ and $1$. This transformation allows to re-write condition~\eqref{eq:rhop} as
\begin{equation}
z^3-\frac{1}{2}z^2+\frac{\nu}{54}=0
\end{equation}
depending on a single parameter $\nu$. Now we substitute $z=t+\frac{1}{6}$ to get the normal form of the cubic equation, 
\begin{equation}
t^3-\frac{1}{12}t+\frac{2\nu-1}{108}=0.
\end{equation}
Let $p=-\frac{1}{12}$ and $q=\frac{2\nu-1}{108}$. If the discriminant $\Delta=q^2+\frac{4}{27}p^2=\frac{\nu(\nu-1)}{54^2}$ is negative, then there 
exist three real solutions. This occurs for $\nu<1$ (i.e., for $\mu<\mu_c$, as expected). In this regime, the three solutions can be expressed 
as~\citep{spiegel:2018}
\begin{equation}
t_k=2\sqrt{\frac{-p}{3}}\cos\left(\frac{1}{3}\cos^{-1}\left(\frac{-q}{2}\sqrt{\frac{27}{-p^3}}\right)+\frac{2\pi k}{3}\right)
\end{equation}
for $k=0,1,2$. Substituting $p=-\frac{1}{12}$ and $q=\frac{2\nu-1}{108}$ and simplifying the resulting expression for $z_k=t_k+\frac{1}{6}$ we obtain: 
for $k=2$,
\begin{equation}
z_2=\frac{1}{6}\left[1-2\sin\left(\frac{1}{3}\sin^{-1}\left(1-2\nu\right)\right)\right];
\end{equation}
and for $k=0$, 
\begin{equation}
z_0=\frac{1}{6}\left[1+2\cos\left(\frac{1}{3}\cos^{-1}\left(1-2\nu\right)\right)\right].
\end{equation}
These two expressions reduce to~\eqref{eq:x12} after going back to the original variables, according to the definitions $x_1:=(\sigma-1)z_2$, 
$x_2:=(\sigma-1)z_0$ and $\nu=\mu/\mu_c$. It can be easily checked that the solution for $k=1$ is negative for all $\nu<1$ and we disregard it.

\section{Analytical bounds to deterministic collapse and recovery times}
\label{sec:appB}
\setcounter{equation}{0}

Consider the cubic equation~\eqref{eq:cubic}. Expressions~\eqref{eq:x12} are explicit solutions for the cubic equation at the values $\rho=\rho_{1,2}$.
The roots $x_{1,2}$ are double and this can be used to obtain the remaining root $x_e$ by simply factorizing the polynomial. The result for $x_e$, after
simplification, is
\begin{equation}\label{eq:x3r}
x_e = \frac{2(\sigma-1)}{3}\left[1+\sin\left(\frac{1}{3}\sin^{-1}\left(1-\frac{2\mu}{\mu_c}\right)\right)\right]
\end{equation}
for $\rho=\rho_1$, and
\begin{equation}\label{eq:x3c}
x_e = \frac{2(\sigma-1)}{3}\left[1-\cos\left(\frac{1}{3}\cos^{-1}\left(1-\frac{2\mu}{\mu_c}\right)\right)\right]
\end{equation}
for $\rho=\rho_2$. We now perturb the cubic equation by changing $\rho_{1,2}$ to $\rho_{1,2}\mp\Delta\rho$, which causes the solutions to vary: in 
particular, a single real solution remains after the change, so we expect that the double root bifurcates into two complex roots, and the root $x_e$ 
remains real. Our aim is to obtain series expansions for the modified solutions for small perturbations $\Delta\rho$.

We focus on the collapse case $\rho=\rho_2$, the expansions for $\rho=\rho_1$ are equivalent. For $\rho_2'=\rho_2+\Delta\rho$ we write 
$x_2'=x_2+\lambda (\Delta\rho)^a$, and our goal is to find the coefficient $\lambda$ and the exponent $a$. Inserting $x_2'$ into Eq.~\eqref{eq:cubic}
and expanding terms we get
\begin{multline}\label{eq:expand}
\lambda\left[3x_2^2-2(\sigma-1)x_2+\sigma(\rho_2-\mu-1)\right](\Delta\rho)^a\\
+\lambda\sigma(\Delta\rho)^{a+1}
+\lambda^2\left[3x_2-(\sigma-1)\right](\Delta\rho)^{2a}\\
+\lambda^3(\Delta\rho)^{3a}+\sigma x_2\Delta\rho=0.
\end{multline}
Notice now that $\rho_2$ satisfies~\eqref{eq:rho12} in terms of $x_2$. Substituting this condition into the first term above we find
\begin{multline}
\lambda\sigma(\Delta\rho)^{a+1}+\lambda^2\left[3x_2-(\sigma-1)\right](\Delta\rho)^{2a}\\+\lambda^3(\Delta\rho)^{3a}+\sigma x_2\Delta\rho=0
\end{multline}
because the first term in~\eqref{eq:expand} turns out to be proportional to $-2x_2^3+(\sigma-1)x_2^2-\mu\sigma$, which vanishes according 
to~\eqref{eq:rhop}. Then the lowest order in $\Delta\rho$ is found by setting $2a=1$, i.e., $a=1/2$. The last expression reduces to
\begin{equation}
\left[\lambda^2\left(3x_2-\sigma+1\right)+\sigma x_2\right]\Delta\rho+\lambda(\lambda^2+\sigma)(\Delta\rho)^{3/2}=0.
\end{equation}
Forcing the leading term to vanish we obtain the coefficient $\lambda$ and the expansion of $x_2'$ up to first order in $\Delta\rho$,
\begin{equation}
x_2'\approx x_2\pm i\sqrt{\frac{\sigma x_2}{3x_2^2-\sigma+1}}(\Delta\rho)^{1/2}.
\end{equation}
It is easy to check that $3x_2^2-\sigma+1\ge0$ using Eq.~\eqref{eq:x12}. Hence the double root $x_2$ splits into two complex conjugate roots
for $\Delta\rho>0$, as expected. According to the notation of the main text, Eq.~\eqref{eq:complex}, we identify $\gamma\approx x_2$ and 
simplifying the imaginary part in terms of the ratio $\nu=\mu/\mu_c$, we find the approximation
\begin{equation}\label{eq:betacol}
\beta\approx \sqrt{\left(\frac{\sigma\Delta\rho}{3}\right)\frac{2\cos\left(\frac{1}{3}\cos^{-1}\left(1-2\nu\right)\right)+1}
{2\cos\left(\frac{1}{3}\cos^{-1}\left(1-2\nu\right)\right)-1}}.
\end{equation}
In a recovery setting ($\rho'=\rho_1-\Delta\rho$), if we repeat \emph{verbatim} the steps followed above for the case of collapse defining 
$x_1'=x_1+\lambda(\Delta\rho)^a$, we obtain the leading-order approximation $x_1'= \gamma\pm i\beta$ with $\gamma\approx x_1$ and 
\begin{equation}\label{eq:betarec}
\beta\approx \sqrt{\left(\frac{\sigma\Delta\rho}{3}\right)\frac{1-2\sin\left(\frac{1}{3}\sin^{-1}\left(1-2\nu\right)\right)}
{1+2\sin\left(\frac{1}{3}\sin^{-1}\left(1-2\nu\right)\right)}},
\end{equation}
respectively.

Next we calculate the first-order correction to $x_e$ in a collapse scenario, $\rho'=\rho_2+\Delta\rho$. In this case we expect that the third root
[given by Eq.~\eqref{eq:x3c}] changes smoothly to $x_e'=x_e+\lambda(\Delta\rho)^a$, with both coefficient $\lambda$ and exponent $a$ to be
determined. Note that expansion~\eqref{eq:expand} remains valid,
\begin{multline}
\lambda\left[3x_e^2-2(\sigma-1)x_e+\sigma(\rho_2-\mu-1)\right](\Delta\rho)^a\\
+\lambda\sigma(\Delta\rho)^{a+1}+\lambda^2\left[3x_e-(\sigma-1\right)](\Delta\rho)^{2a}+\\
\lambda^3(\Delta\rho)^{3a}+\sigma x_e\Delta\rho=0,
\end{multline}
but now the first term does not cancel because $x_e$ is not a relative extremum of $\rho(x)$. The leading order is therefore obtained for $a=1$ and
$\lambda$ has to satisfy the condition
\begin{equation}
\lambda\left[3x_e^2-2(\sigma-1)x_e+\sigma(\rho_2-\mu-1)\right]+\sigma x_e=0.
\end{equation}
From this we can solve for $\lambda$ and get the approximated real root
\begin{equation}
x_e'\approx x_e-\frac{\sigma x_e\Delta\rho}{3x_e^2-2(\sigma-1)x_e+\sigma(\rho_2-\mu-1)}
\end{equation}
up to leading order in $\Delta\rho$. Simplifying we finally get the third root expressed as
\begin{equation}\label{eq:xecol}
x_e'\approx x_e+\frac{8}{3}\left(\frac{\sigma\Delta\rho}{\sigma-1}\right)\frac{\cos\left(\frac{1}{3}\cos^{-1}\left(1-2\nu\right)\right)-1}
{\left[2\cos\left(\frac{1}{3}\cos^{-1}\left(1-2\nu\right)\right)-1\right]^2}
\end{equation}
with $x_e$ given by Eq.~\eqref{eq:x3c} for the case of collapse and, repeating the exact same analysis for a recovery situation, we find
\begin{equation}\label{eq:xerec}
x_e'\approx x_e+\frac{8}{3}\left(\frac{\sigma\Delta\rho}{\sigma-1}\right)\frac{\sin\left(\frac{1}{3}\sin^{-1}\left(1-2\nu\right)\right)+1}
{\left[2\sin\left(\frac{1}{3}\sin^{-1}\left(1-2\nu\right)\right)+1\right]^2}
\end{equation}
where now $x_e$ is given by~\eqref{eq:x3r}.

Equipped with these approximations, we are now able to derive the leading terms of collapse and recovery times in their expansion in powers
of $\Delta\rho$, cf. Eqs.~\eqref{eq:colapp} and~\eqref{eq:recapp}. Let us focus on Eqs.~\eqref{eq:T} and~\eqref{eq:collapse} for collapse times.
Note first that the dominant, diverging term as $\Delta\rho\to 0$ in $\tau(x;\vv{p})$ is precisely the one with the arctangent: the factor $\beta^{-1}$
yields a $(\Delta\rho)^{-1/2}$ divergence as $\Delta\rho$ comes closer to zero, which dominates any logarithmic divergence that can arise from the
second term in $\tau(x;\vv{p})$. Therefore the leading term of the collapse time, in the limit $\Delta\rho\to 0$, is
\begin{multline}\label{eq:aux}
T_{col}\approx \frac{\sigma(\gamma+1)}{\beta(x_e'-\gamma)}\left[\tan^{-1}\left(\frac{x_e'+\Delta x-\gamma}{\beta}\right)\right.\\
\left.-
\tan^{-1}\left(\frac{x_2-\gamma}{\beta}\right)\right]+O(\log\Delta\rho).
\end{multline}
As we have shown above, $\gamma\approx x_2+O(\Delta\rho)$, hence $\frac{x_2-\gamma}{\beta}\sim (\Delta\rho)^{1/2}$ and
$\beta^{-1}\tan^{-1}\left(\frac{x_2-\gamma}{\beta}\right)\sim O(1)$. However, $x_e'+\Delta x-\gamma$ is non-zero in the limit, hence the first 
arctangent in~\eqref{eq:aux} dominates (diverges) as $\Delta\rho\to 0$. Approximating the arctangent by $\pi/2$ in the limit $\Delta\rho\to 0$ we 
finally obtain
\begin{equation}\label{eq:tapp}
T_{col}\approx \frac{\pi\sigma(\gamma+1)}{2\beta(x_e'-\gamma)}.
\end{equation}
Substituting of $\gamma\approx x_2$, using Eqs.~\eqref{eq:betacol}, \eqref{eq:xecol} and~\eqref{eq:x12} and simplifying the result we obtain
the lower bound for collapse times given by Eq.~\eqref{eq:colapp}. Because the arguments leading to~\eqref{eq:tapp} hold both for collapse and
recovery times, we can estimate as well
\begin{equation}
T_{rec}\approx \frac{\pi\sigma(\gamma+1)}{2\beta(x_e'-\gamma)}
\end{equation}
where $\gamma\approx x_1$ and $\beta$, $x_e'$ and $x_1$ are given by Eqs.~\eqref{eq:betarec}, \eqref{eq:xerec} and~\eqref{eq:x12}, respectively.
After simplification we obtain precisely our approximation~\eqref{eq:recapp} for the time to recover after crossing the threshold $\rho_1$.

\section{Deterministic limit of mean first-passage times}
\label{sec:appC}
\setcounter{equation}{0}

In this section we show that the deterministic dynamics~\eqref{eq:imscal} can be recovered from the stochastic birth-death process in the limit
$b\to 0$. Consider the recurrence relation given by Eq.~\eqref{eq:fptrec} satisfied by mean first-passage times. Multiplying by $b/r$, this relation can be 
written in terms of the scaled time $\bar{\tau}=r\bar{t}$ and two $z$-abundances, namely $z_0=bn_0$ and $z_1=bn_1$, as
\begin{multline}\label{eq:fptcont}
q^+(z)[\bar{\tau}(z,z_1)-\bar{\tau}(z+b,z_1)]\\
+q^-(z)[\bar{\tau}(z,z_1)-\bar{\tau}(z-b,z_1)]=b.
\end{multline}
Here we have omitted the $0$ subindex and $z$ stands for the initial state used to calculate the mean first-passage time. We assume that 
$z<z_1$ (the case $z>z_1$ can be straightforwardly analyzed in a similar way). Now expand $\bar{\tau}(z\pm b,z_1)$ about $z$ up to second 
order in $b$,
\begin{equation}
\bar{\tau}(z\pm b,z_1)=\bar{\tau}(z,z_1)\pm b\frac{\partial }{\partial z}\bar{\tau}(z,z_1)
+\frac{b^2}{2}\frac{\partial^2}{\partial z^2}\bar{\tau}(z,z_1)+\dots,
\end{equation}
hence equation~\eqref{eq:fptcont} re-writes as 
\begin{multline}\label{eq:ODEtau}
[q^-(z)-q^+(z)]\frac{\partial }{\partial z}\bar{\tau}(z,z_1)\\
-
\frac{b}{2}[q^-(z)+q^+(z)]\frac{\partial^2}{\partial z^2}\bar{\tau}(z,z_1)=1-O(b^2).
\end{multline}
This is an ordinary differential equation which depends parametrically on $z_1$, i.e., the final endstate is fixed. In the limit $b\to 0$ solving this ODE 
amounts to integrating
\begin{equation}
[q^-(z)-q^+(z)]\frac{\partial }{\partial z}\bar{\tau}(z,z_1)=1,
\end{equation}
i.e.,
\begin{equation}\label{eq:b0}
\frac{\partial }{\partial z}\bar{\tau}(z,z_1)=\frac{1}{q^-(z)-q^+(z)}.
\end{equation}
It is important to remark that the partial derivative is independent of the endstate $z_1$. Because $z<z_1$, integration yields
\begin{equation}\label{eq:b0int}
\bar{\tau}(z,z_1)=\int_{z}^{z_1}\frac{d\zeta}{q^-(\zeta)-q^+(\zeta)}+\bar{\tau}_1(z_1)
\end{equation}
for an unknown function $\bar{\tau}_1$.

Now observe that the time to go from $z$ to an arbitrary endstate $z_1$ is $\bar{\tau}(z,z_1)$, whereas the time to go from $z+\Delta z$ 
to the \emph{same endstate} $z_1$ is $\bar{\tau}(z+\Delta z,z_1)$ for any arbitrary increment $\Delta z$. Therefore, applying additivity of 
integration to~\eqref{eq:b0int}, if $\Delta z>0$ it holds that the time to go from $z$ to $z+\Delta z$ is
\begin{equation}
\bar{\tau}(z,z+\Delta z)=\bar{\tau}(z,z_1)-\bar{\tau}(z+\Delta z,z_1).
\end{equation}
Expanding the rhs in power series for infinitesimally small $\Delta z$, we express the time to go from 
$z$ to $z+\Delta z$ as
\begin{multline}
\bar{\tau}(z,z+\Delta z)\approx-\frac{\partial }{\partial z}\bar{\tau}(z,z_1)\Delta z\\
=
\frac{\Delta z}{q^+(z)-q^-(z)}=\frac{\Delta z}{z\left(1-\frac{z}{\sigma}\right)-\frac{\rho\, z}{1+z}+\mu},
\end{multline}
where we have used~\eqref{eq:b0} and~\eqref{eq:qpm}. This yields, in the limit $\Delta z\to 0$, the differential equation
\begin{equation}
\frac{d\bar{\tau}}{dz}=\left[z\left(1-\frac{z}{\sigma}\right)-\frac{\rho z}{1+z}+\mu\right]^{-1},
\end{equation}
which is precisely the deterministic differential equation~\eqref{eq:imscal}. Therefore, mean first-passage times converge to deterministic
ones as carrying capacity $K$ grows to infinity for fixed $\sigma$. This phenomena was observed numerically in Fig.~\ref{fig:times} for collapse
and recovery times.

A final remark is on purpose here. We have analyzed the differential equation~\eqref{eq:ODEtau} in the case $b=0$, i.e., in the deterministic limit.
However, such an expansion can be used to construct analytical approximations for the times to collapse and recovery for \emph{finite} $b$, similar
to those we have obtained for the deterministic model, which are based on convenient approximations to integrals of rational functions. This would help
understand how stochastic collapse and recovery times depend on model parameters. We leave the analysis of these approximations for future 
extensions of our work.

%% not want to use model2-names.bst.

\bibliographystyle{model2-names}
\bibliography{pescaos}

\end{document}